\documentclass[conference]{IEEEtran}
\IEEEoverridecommandlockouts
\usepackage{cite}
\usepackage{amsmath,amssymb,amsfonts, mathtools, comment}
\usepackage{tablefootnote}
\usepackage{algorithmic}
\usepackage{graphicx}
\usepackage{textcomp}
\usepackage{xcolor}
\usepackage[ruled,lined,linesnumbered]{algorithm2e}
\def\BibTeX{{\rm B\kern-.05em{\sc i\kern-.025em b}\kern-.08em
    T\kern-.1667em\lower.7ex\hbox{E}\kern-.125emX}}
\usepackage{orcidlink}
\SetKwComment{Comment}{//~}{}
\SetCommentSty{normalfont}
\SetNlSty{}{}{:}

\usepackage{xcolor}
\definecolor{myBlue}{rgb}{0.0,0.0,1.0}
\definecolor{myRed}{rgb}{1.0, 0.0, 0.0}
\definecolor{myGreen}{rgb}{0.0, 0.6, 0.0}

\begin{document}

\title{MX-SAFE: Versatile Inference- and Training-Proof Microscaling Format with On-the-Fly Exponent and Mantissa Bit Allocation}


\author{\IEEEauthorblockN{Dahoon Park\IEEEauthorrefmark{1}\IEEEauthorrefmark{2}\orcidlink{0000-0002-8047-2416}, Jahyun Koo\IEEEauthorrefmark{1}\IEEEauthorrefmark{3}\orcidlink{0000-0001-5652-5306}, Sangwoo Hwang\IEEEauthorrefmark{2}\orcidlink{0000-0001-8716-5390}, and Jaeha Kung\IEEEauthorrefmark{4}\IEEEauthorrefmark{2}\orcidlink{0000-0001-6151-8602}}
\IEEEauthorblockA{$\dagger$~\textit{School of Electrical Engineering, Korea University, South Korea}}
\IEEEauthorblockA{$\ddagger$~\textit{Department of Electrical Engineering and Computer Science, DGIST, South Korea}}
\vspace{-5mm}
\thanks{This work was partially supported by the Institute of Information \& Communications Technology Planning \& Evaluation (IITP) funded by the Korea government (MSIT) under Grant RS-2023-00229849 and Grant RS-2025-25442405; in part by the National Research Foundation of Korea (NRF) funded by the Ministry of Science and ICT under Grant NRF-2023R1A2C2006290. The EDA tool was supported by the IC Design Education Center (IDEC) in South Korea.}
\thanks{\IEEEauthorrefmark{1}\hspace{0.5mm} D. Park and J. Koo are equally contributed authors.}
\thanks{\IEEEauthorrefmark{4}\hspace{0.5mm} J. Kung is the corresponding author (email: \textit{jhkung@korea.ac.kr}).}
}

\maketitle



\begin{abstract}
As the demand for deep learning grows, cost reduction through quantization has become essential for both training and inference. In 2022, the Open Compute Project (OCP) consortium standardized narrow precision formats for deep learning, called the microscaling (MX) format. 
The MX format is a hardware-friendly dynamic quantization scheme that effectively reduces the data size by sharing an 8-bit exponent across multiple operands. 
The MX format can be categorized into two types with their own strengths: (i) MXINT which focuses on a high precision consisting only of mantissa bits and (ii) MXFP which focuses on a wider dynamic range by allowing local exponent bits. 
In this work, we present a versatile MXFP format, called MX-SAFE (MXSF in short), that adaptively uses two modes, i.e., a wider mantissa mode (FP8\_E2M5) and a subnormal FP mode (FP5\_E3M2), to support both training and direct-cast inference.
Furthermore, we propose a tile-based block design to increase hardware efficiency by reducing the burden of re-quantization process during the training with the MXSF format.
Owing to the use of the proposed MXSF format, 0.05\%/11.1\% and 3.55\%/3.57\% improvements in accuracy, on average, for inference/full-training compared to MXFP8\_E2M5 and MXFP8\_E4M3 are observed, respectively. 
Moreover, we present a training-inference accelerator that supports the MXSF format and it achieves similar accuracy to the BF16 baseline while using 24.9\% less total energy consumption.

\end{abstract}


\section{Introduction}
As the size of deep neural networks continues to grow, quantization has become one of the most effective compression techniques.
The quantization technique can be broadly categorized into static and dynamic quantization.
The static quantization has the advantage of being able to compress inputs and weights based on pre-computed statistics, enabling efficient computation and reduced memory requirement.
However, as recent models have become more complex, some intermediate operations require dequantization process to floating point representation (e.g., softmax and layer norm), and each model has its own optimal statistical data.
Particularly, weight parameters keep changing during training which makes it very challenging to use the static quantization.
The dynamic quantization, on the other hand, extracts tensor statistics on the fly by inspecting the incoming data and then performs quantization. 
This allows for more accurate quantization~\cite{spinquant, so2024temporal}, but it comes at the cost of running quantization-dequantization steps.

\begin{figure}[t]
    \centering
    \includegraphics[width=0.95\linewidth]{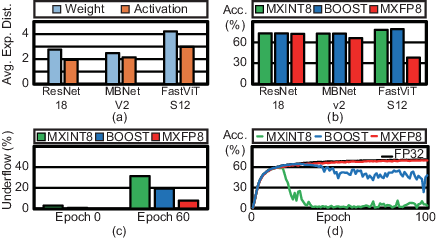}
    \caption{Impact of various MX formats on inference (a-b) and training (c-d): (a) average distance to the  max exponent within a block during inference, (b) direct-cast inference accuracy on ImageNet-1K, (c) ratio of weight gradient underflows at FastViT-T8's \textit{stages.0.blocks.0.mlp.fc1} layer during training, and (d) training curve for FastViT-T8.}\vspace{-4mm}
    \label{fig:fig1_intro}
\end{figure}

Recently, microscaling (MX) format has been standardized as a new data format by Open Compute Project~\cite{mx} which simplifies the dynamic quantization process, efficiently converting \texttt{FP32} computations to low-bit computations. 
The MX format is currently supported by commodity deep learning hardware such as NVIDIA Blackwell Tensor Cores~\cite{nvidiablackwell}, Microsoft MAIA100~\cite{msmaia100}, and Tenstorrent Wormhole~\cite{tens_wormhole}, as well as custom accelerators~\cite{opal, bsfp, boost, lee2023dbps, zhang2022fast}.
With various MX formats being proposed, however, it is difficult for us to select the best MX format which varies depending on the task, either inference or training.
For direct-cast inference\footnote{Directly casting \texttt{FP32} elements to lower bit-width representation, e.g., \texttt{MXINT8} or \texttt{MXFP8}, without any training and calibration.}, the distance between the max exponent and each element's exponent within a block is small (Fig.~\ref{fig:fig1_intro}(a)), making the mantissa bit-width critical.
Consequently, \texttt{MXINT8} and BOOST (\texttt{MXFP8\_E2M5})~\cite{boost} outperform \texttt{MXFP8\_E4M3} (Fig.~\ref{fig:fig1_intro}(b)).
However, as shown in Fig.~\ref{fig:fig1_intro}(c), a significant portion of small gradient values become zeros (i.e., underflows) for MX formats with limited local exponent bits.
Thus, only \texttt{MXFP8\_E4M3} format which can represent small values using local exponent bits, on top of the shared exponent, shows the similar training curve to \texttt{FP32} while other MX formats fail to converge (Fig.~\ref{fig:fig1_intro}(d)).

These motivational experiments on both inference and training imply that each MX format type has its own benefits (\texttt{MXINT} and \texttt{BOOST} for a higher precision and \texttt{MXFP} for a wider dynamic range), but no universal solution exists.
Therefore, for the first time, we present a versatile 8-bit \texttt{MXFP} format, named \texttt{MXSF}, 
that guarantees safe direct-cast inference and full training. 
Our \texttt{MXSF} is defined based on the quantitative error analysis, so that we can switch between \texttt{MXFP} formats with wide mantissas and wide exponents on the fly.
To test the validity of the proposed \texttt{MXSF}, a wide range of deep learning models, from CNNs to transformers, and both inference and training tasks are evaluated.

Our main contributions can be summarized as follows:
\begin{itemize}
    \item \textbf{Quantitative analysis on MXINT/MXFP}: We quantify the quantization error that occurs during conversion from \texttt{FP32}/\texttt{BF16} to \texttt{MXINT} and \texttt{MXFP}. 
    Our analysis shows that the optimal MX format changes with respect to the exponent distance to the shared exponent within a block. 
    \item \textbf{Safe MX format}: We propose a novel \texttt{MX-SAFE} format, i.e., \texttt{MXSF} in short, which safely supports direct-cast inference and full-training/fine-tuning.
    The \texttt{MXSF} format dynamically stores an element with \texttt{MXFP8\_E2M5} or with \texttt{MXFP5\_E3M2} that targets to reduce both quantization errors and underflows.
     \item \textbf{MX-SAFE hardware accelerator}: We design a MAC unit that handles \texttt{MXSF} formats and implement a systolic tensor array built out of the proposed \texttt{MXSF}-aware MACs.
     

\end{itemize}


\section{Backgrounds}
\subsection{Microscaling (MX) Data Format}
Recently, microscaling (MX) data formats~\cite{mx} have been presented showing their effectiveness in direct-cast inference or training of various deep learning models. 
The original MX format packs 32 data into a block (i.e., block\_size = 32), shares one exponent per block ($S_e$), and aligns each element in the block according to the shared exponent.
While sharing the global exponent, each element in the block can be represented by either an integer format, e.g., \texttt{MXINT8}, or a floating point format, e.g., \texttt{MXFP8\_E4M3} or \texttt{MXFP8\_E5M2}\footnote{Here, `\texttt{E}' and `\texttt{M}' denote (local) exponent bits and mantissa bits, respectively. The total bit-width $N$ becomes $E+M+1$ where 1-bit is for sign.}. 
One may also choose to use a lower-bit floating point representation, e.g., \texttt{MXFP4\_E2M1} or \texttt{MXFP6\_E2M3}, \texttt{MXFP6\_E3M2}.

\subsection{Optimal MX Format for Inference \& Training}
While the MX format has many different variants, there is an ambiguity in that the optimal type/format depends on the task at hand.
During inference, tensor distributions show relatively smaller variances compared to gradients in training.
In other words, local exponents within a block are close to each other making \texttt{MXINT8} and \texttt{MXFP8\_E2M5} represent the tensors more accurately compared to \texttt{MXFP8\_E4M3} owing to wider mantissa bits. 
On the other hand, during training, gradients have high variance, leading to larger differences between local exponent values within a block, which causes many underflows if exponent bits are not properly assigned.
Thus, the model trained in \texttt{MXFP8} with wider exponent bits outperforms the one trained in \texttt{MXINT8}.
However, supporting both INT and FP types in hardware introduces high overheads. 
While there are some prior works demonstrating training with \texttt{MXFP4\_E2M1}~\cite{tetrajet, amazon_mxfp4}, they have notable drawbacks. 
First, \cite{tetrajet}-based training results in suboptimal training performance in CNNs (Section~\ref{sec:analysis_required_bit}).
Furthermore, the method presented in~\cite{amazon_mxfp4} requires additional hardware blocks for tensor rotation which incurs latency overhead~\cite{lightrot}.
Also, both works perform \texttt{BF16} computations on $Q\cdot\!K^T$ and $Attn\cdot\!V$ in attention layers.
Thus, in this work, we \textit{stick to an 8-bit MX format for all computations} and \textit{develop a new MX format} that appropriately combines advantages of both MX data types with minimal hardware overhead.

\section{Quantitative Analysis on MX Formats}
\subsection{Analytical Comparison Between MXINT and MXFP}\label{sec:err_analysis}

Prior to proposing a new MX data format, we start by analytically comparing \texttt{MXINT} and \texttt{MXFP} formats.
The following equation provides the quantized value in \texttt{MXINT} for a given floating point value $x$.
\begin{equation}\label{eq:mxint}
\begin{split}
    MXINT(x)=s_x\cdot 2^{S_e}\cdot\left\lfloor\frac{1.m_x \cdot 2^{(m_i-2)}}{2^{(S_e-e_x)}}\right\rceil \cdot 2^{-(m_i-2)},
\end{split}
\end{equation}
where $x=s_x\cdot 2^{e_x} \cdot 1.m_x$ in \texttt{FP32}/\texttt{BF16}, $s_x$ is the sign, $e_x$ is the exponent, and $m_x$ is the mantissa value of $x$.
The $S_e$ is the shared exponent within the block containing $x$ and $m_i$ is the \texttt{MXINT}'s mantissa bit-width.
On the other hand, the same value $x$ can be represented by the \texttt{MXFP} format as follows:
\begin{equation}
    MXFP(x) = \begin{cases}
        FP_{norm}(x) & \text{if $x_{le} > 0$}, \\
        FP_{subnorm}(x) & \text{if $x_{le} \leq 0$}, 
    \end{cases}
\end{equation}
where the local exponent $x_{le} = E - (S_e - e_x)$.
Here, $E$ is the maximum local exponent value, i.e., $2^{e_f} - 1$, where $e_f$ is the \texttt{MXFP}'s local exponent bits.
The $FP_{norm}$ and $FP_{subnorm}$ are defined as follows:
\begin{gather}\label{eq:mxfp}
    FP_{norm}(x) = s_x\cdot 2^{S_e-E}\cdot 2^{x_{le}}\cdot \Bigl\lfloor1.m_x \cdot 2^{m_f}\Bigr\rceil \cdot 2^{-m_f}, \\
    FP_{subnorm}(x) = s_x\cdot 2^{S_e-E}\cdot \Bigl\lfloor0.m_x \cdot 2^{m_f + x_{le}}\Bigr\rceil \cdot 2^{-m_f},
\end{gather}
where $m_f$ is the \texttt{MXFP}'s mantissa bits.

Based on Eq.~(\ref{eq:mxint}) and Eq.~(3-4), the maximum quantization errors of \texttt{MXINT} and \texttt{MXFP} formats can be formulated by
\begin{align}
    \Delta_{MXINT}&=2^{S_e-(m_i-2)}\cdot 2^{(S_e-e_x)-(m_i-2)},\label{eq:int_err} \\ \Delta_{MXFP}&=2^{e_x-m_f}\cdot 2^{-min(x_{le}, 0) -m_f}. \label{eq:fp_err}
\end{align}
For both cases, the error occurs from the round function which is the second multiplicative term in Eq.~(\ref{eq:int_err}-\ref{eq:fp_err}).
In case of \texttt{MXINT}, error occurs due to the difference between the shared exponent $S_e$ and the exponent of $x$ ($e_x$), as well as the supported mantissa bits ($m_i$).
In contrast, in \texttt{MXFP}, it has the local exponent $x_{le}$, and as long as $x_{le}$ does not fall below zero, the rounding error is only caused by the supported mantissa bits ($m_f$).
Based on Eq.~(\ref{eq:int_err}-\ref{eq:fp_err}), we can compare \texttt{MXINT8} and \texttt{MXFP8\_E2M5} in terms of quantization error at various `$S_e-e_x$' distances.
Only when $S_e-e_x=0$, \texttt{MXINT8} provides a smaller error than \texttt{MXFP8\_E2M5}.
If $S_e-e_x=1$, both experiences the same quantization error.
When $S_e-e_x$ exceeds 1, then \texttt{MXFP8\_E2M5} introduces a smaller quantization error.
Since both weight and activation tensors exhibit the average exponent distance ($S_e-e_x$) greater than 2 (Fig.~\ref{fig:fig1_intro}(a)),
\texttt{MXFP8\_E2M5} is expected to have a lower quantization error than \texttt{MXINT8}, as summarized in Table~\ref{tab:model_error}.
Therefore, we establish \texttt{MXFP8\_E2M5} as the baseline for inference. 
The following section will discuss how this \texttt{MXFP8} format can be modified to make it suitable for training.

\begin{table}[t]
\caption{Mean Squared Errors of Directly Casting \\into Different MX Formats in Various Models}
\label{tab:model_error}
\centering\resizebox{\linewidth}{!}{
\begin{tabular}{|l|cc|cc|cc|}
\hline
\multicolumn{1}{|c|}{\textbf{Model}} & \multicolumn{2}{c|}{\textbf{ResNet-18}}                                                                                              & \multicolumn{2}{c|}{\textbf{MobileNetV2}}                                                                                            & \multicolumn{2}{c|}{\textbf{FastViT-T8}}                                                                                            \\ \hline \hline
\multicolumn{1}{|c|}{Format / Type}  & \multicolumn{1}{c|}{\begin{tabular}[c]{@{}c@{}}Act.\\ $(10^{\text{-}4})$\end{tabular}} & \begin{tabular}[c]{@{}c@{}}Weight\\ $(10^{\text{-}6})$\end{tabular} & \multicolumn{1}{c|}{\begin{tabular}[c]{@{}c@{}}Act.\\ $(10^{\text{-}4})$\end{tabular}} & \begin{tabular}[c]{@{}c@{}}Weight\\ $(10^{\text{-}6})$\end{tabular} & \multicolumn{1}{c|}{\begin{tabular}[c]{@{}c@{}}Act.\\ $(10^{\text{-}4})$\end{tabular}} & \begin{tabular}[c]{@{}c@{}}Weight\\ $(10^{\text{-}4})$\end{tabular} \\ \hline
\texttt{MXINT8}                               & \multicolumn{1}{c|}{\color[HTML]{6434FC}\textbf{1.31}}                                                  & 6.13                                                    & \multicolumn{1}{c|}{11.2}                                                  & 3.41                                                    & \multicolumn{1}{c|}{1.27}                                                  & 1.48                                                    \\ \hline
\texttt{MXFP8\_E2M5}                          & \multicolumn{1}{c|}{\color[HTML]{FE0000}\textbf{1.14}}                                                  & \color[HTML]{FE0000}\textbf{3.41}                                                     & \multicolumn{1}{c|}{\color[HTML]{FE0000}\textbf{8.35}}                                                   & \color[HTML]{FE0000}\textbf{2.29}                                                     & \multicolumn{1}{c|}{\color[HTML]{FE0000}\textbf{0.79}}                                                  & \color[HTML]{FE0000}\textbf{1.00}                                                     \\ \hline
\texttt{MXFP8\_E4M3}                          & \multicolumn{1}{c|}{17.3}                                                  & 42.5                                                    & \multicolumn{1}{c|}{120}                                                   & 31.0                                                    & \multicolumn{1}{c|}{10.3}                                                  & 13.1                                                    \\ \hline
\texttt{MXSF} (Proposed)                          & \multicolumn{1}{c|}{1.40}                                                  & \color[HTML]{6434FC}\textbf{5.07}                                                     & \multicolumn{1}{c|}{\color[HTML]{6434FC}\textbf{10.7}}                                                   & \color[HTML]{6434FC}\textbf{3.06}                                                     & \multicolumn{1}{c|}{\color[HTML]{6434FC}\textbf{1.10}}                                                  & \color[HTML]{6434FC}\textbf{{1.17}}                                                     \\ \hline
\multicolumn{7}{l}{\scriptsize * Best and second best are marked in {\color[HTML]{FE0000}red} and {\color[HTML]{6434FC}blue}, respectively.}\vspace{-2mm} \\
\end{tabular}}\end{table}

\begin{figure}[t]
    \centering
    \includegraphics[width=0.9\linewidth]{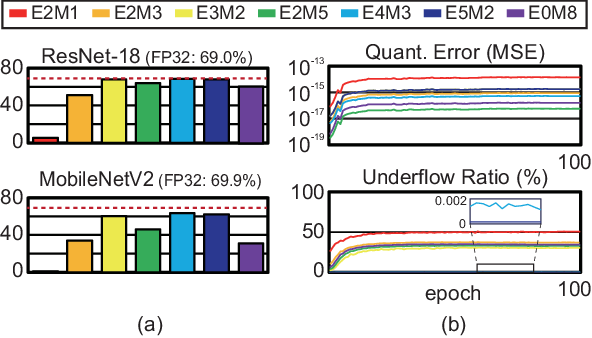}\vspace{-4mm}
    \caption{(a) Training performance with various microscaling formats. The \textit{red dotted line} represents FP32 training accuracy. (b) Quantization error and underflow ratio during training (target: ResNet-18's \textit{layer2.1.conv2.grad})}\vspace{-4mm}
    \label{fig:fig2_training_compare}
\end{figure}

\subsection{Analysis on Required Bit-precision in Training}\label{sec:analysis_required_bit}

According to Table~\ref{tab:model_error},  \texttt{MXINT8} and \texttt{MXFP8\_E2M5} have lower quantization error than \texttt{MXFP8\_E4M3}.
However, those two MX formats paradoxically exhibit training instability as shown in Fig.~\ref{fig:fig1_intro}(d).
This is mainly because the occurrence of underflows has a more significant impact on training stability than quantization errors.
Fig.~\ref{fig:fig2_training_compare}(a) illustrates the results of various MX-format training runs\footnote{Here, \texttt{MXFP4} is based on TetraJet~\cite{tetrajet} Q-EMA algorithm.} on ResNet-18~\cite{resnet} and MobileNetV2~\cite{mobilenetv2}.
For both benchmarks, at least three exponent bits were required for stable training.
However, for those MX formats with wider exponent bits, the quantization error increases due to narrower mantissa bits, which results in a noticeable accuracy degradation in MobileNetV2.

To investigate the cause of training instability in detail, we analyzed the quantization error and underflow ratio of activation gradients at the \textit{layer2.1.conv2} layer in ResNet-18. 
The analysis was performed by converting the gradients into various MX formats and monitoring the metrics across training epochs.
The results shown in Fig.~\ref{fig:fig2_training_compare}(b) reveal that while \texttt{MXINT8} (E0M8) and \texttt{BOOST} (E2M5) show lower quantization error than \texttt{MXFP8\_E4M3}, they have significantly higher underflow ratios, which prevents the model from stable training. 
This suggests that even gradients with very small magnitudes are critical to training performance and must be preserved.
Based on these analyses, we propose \texttt{MXSF}, \textit{a unified MX format that deals with both underflows and quantization errors} by dynamically adjusting exponent and mantissa bits based on the criteria explained in Section~\ref{sec:err_analysis}.

\begin{figure*}[t!]
    \centering
    \includegraphics[width=0.95\textwidth]{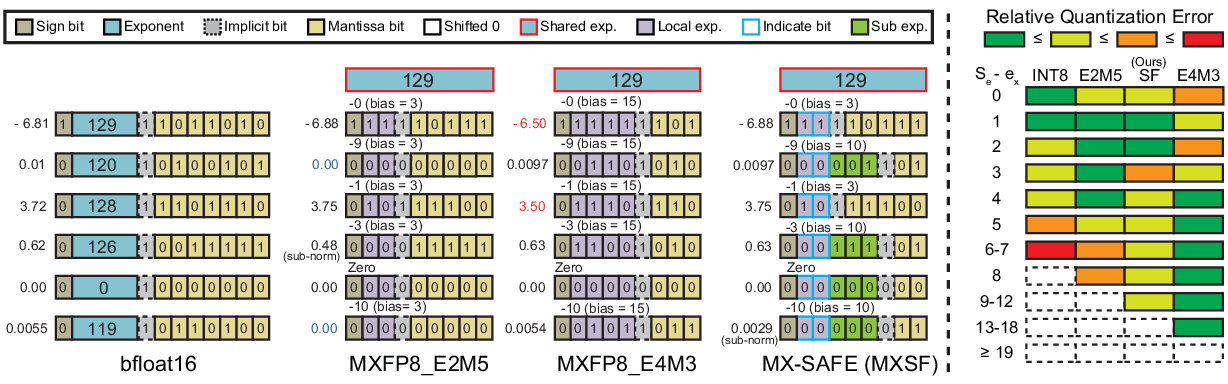}\vspace{-2mm}
    \caption{(Left) \texttt{Bfloat16} as a baseline and various microscaling data formats, i.e., \texttt{MXFP8\_E2M5}, \texttt{MXFP8\_E4M3}, and proposed MX-SAFE (\texttt{MXSF}). The blue values in \texttt{MXFP8\_E2M5} represents the underflow and the red values in \texttt{MXFP8\_E4M3} show conversion errors larger than 0.2. (Right) Visualization of relative quantization error with respect to a distance between the shared exponent and the local exponent ($S_e-e_x$). The blank dashed box denotes underflow.}
    \label{fig:fig3_vis_mxsf_format}\vspace{-2mm}
\end{figure*}

\begin{algorithm}[t!]
\caption{MXSF Conversion}\label{alg:mxif}
\KwData{$X$: floating point numbers within a block} 
\KwResult{$\hat{X}$: converted \texttt{MXSF} data for tensor $X$}
$S_e \gets floor(log_2(max(|X|)))$\; 
$X_e \gets floor(log_2(|X|))$\;
\eIf{$(S_e - X_e)  < 3$}{
    $\hat X \gets MXFP(X, e_f=2, m_f=5, bias=3)$;}{
    $\hat                                 X \gets MXFP(X, e_f=3, m_f=2, bias=10)$;}
\end{algorithm}

\section{MX-SAFE: Versatile Microscaling Format}
\subsection{Proposed MXSF Data Format}
The proposed \texttt{MXSF} format, named after a safe MX format, is designed to support two FP formats in a single block.
The main goal of supporting two data types in a single block is to minimize quantization errors and reduce underflows.
This makes the proposed \texttt{MXSF} versatile, so that we can use this format in any AI deployment stages.
In this work, we focus on designing the \texttt{MXSF} format that supports 
a narrow exponent + large mantissa (\texttt{E2M5}), which shows similar accuracy to \texttt{FP32} for inference, and a wider exponent with extra bias + small mantissa (\texttt{E3M2} + bias), which performs well during training by suppressing underflows.

To accommodate both within a single block, we propose a format that allocates more exponent bits, on the fly, for elements with a large exponent gap (i.e., $S_e-e_x$) as illustrated in Fig.~\ref{fig:fig3_vis_mxsf_format}. 
Specifically, we repurpose the subnormal representation in \texttt{MXFP8\_E2M5}, where local exp $x_{le}= 00$ (the 2\textsuperscript{nd} and the last elements in Fig.~\ref{fig:fig3_vis_mxsf_format}), to better represent small numbers, denoted \texttt{E3M2} (or sub-FP), using the remaining 5 bits.
In the original \texttt{MXFP8\_E2M5} format, values with an exponent gap of 3 or larger ($S_e-e_x\ge3$) fall into the subnormal category. 
In our \texttt{MXSF}, this region is represented by using the sub-FP format, effectively extending the minimum exponent value from -3 down to -9.
The addition of \texttt{E3M2} format extends the representable value range without overlapping the range of \texttt{E2M5} with proper biasing (bias$=10$).
This allows our modified representation to cover a dynamic range similar to that of \texttt{MXFP8\_E3M4}, slightly lower than \texttt{E4M3}.
As shown in Fig.~\ref{fig:fig3_vis_mxsf_format}, when the exponent gap is 2 or less, \texttt{MXSF} behaves same as \texttt{MXFP8\_E2M5}, preserving direct-cast inference accuracy. 
When an element within a block has an exponent gap greater than 2, local exponent becomes `$00$' indicating the use of \texttt{E3M2} dynamically extending the dynamic range for small values, improving the training stability.
The conversion process for \texttt{MXSF} is described in Alg.~\ref{alg:mxif}.

\begin{figure}[t]
    \centering
    \includegraphics[width=0.9\linewidth]{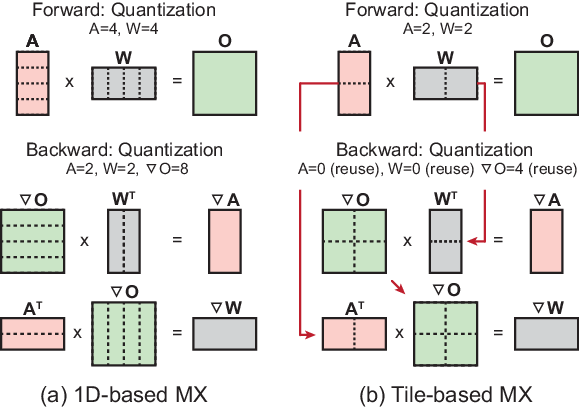}\vspace{-3mm}
    \caption{The forward and backward passes for a linear layer during training. (a) 1D-based MX block (1$\times$4) and (b) 2D tile-based MX block (2$\times$2). The \textit{red arrows} highlight the reuse of quantized MX blocks.}
    \label{fig:tile_quant}\vspace{-2mm}
\end{figure}

\subsection{MX Block Tiling for Inference/Training}

As shown in Fig.~\ref{fig:tile_quant}(a), the commonly used 1D block design in the current MX format is suited at the forward-only computation flow of inference. 
In this setting, quantized weights remain unchanged and activations are not reused, making the 1D block structure suitable without any issues. 
However, during the training process, particularly in the backward pass, both weights and activations are transposed and used for computing gradients.
Due to the transpose operation, weight/activation tensors need to be dequantized and re-quantization is needed along the transposed dimension, which incurs latency overhead.
Fig.~\ref{fig:tile_quant} shows a simple example with the number of MX quantization process for each tensor.

To address this limitation, as illustrated in Fig.~\ref{fig:tile_quant}(b), we propose 2D tile-based MX block design. 
This 2D tiling strategy makes us to maintain the quantized tensor in its original shape and reuse it during the backward pass without additional dequantize/quantize step.
Accordingly, we design a training-inference accelerator using our versatile \texttt{MXSF} that supports both 1D and 2D MX blocks, maximizing overall hardware efficiency.
The 1D block tiling is also supported in the hardware for the inference with batch size of 1.

\begin{figure}[t]
    \centering
    \includegraphics[width=0.85\linewidth]{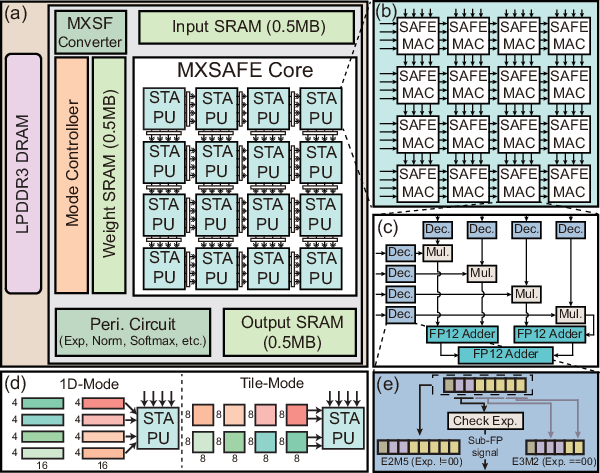}\vspace{-2mm}
    \caption{Overview of MX-SAFE accelerator architecture.}
    \label{fig:MXSF_System}\vspace{-4mm}
\end{figure}

\section{MXSF-based Multi-format Systolic Tensor Array Accelerator}\label{sec:mxsf_architecture}

\subsection{MX-SAFE Accelerator}
Fig.~\ref{fig:MXSF_System}(a) shows overall architecture of the MX-SAFE accelerator, which is based on a systolic tensor array (STA~\cite{liu2020systolic}). 
The STA generalizes the conventional systolic array by grouping multiple MACs to form a computing block, where each MAC contains multiple multipliers (\textit{vectorized}).
Grouping multiple MACs as a block, i.e., processing unit (PU), reduces the number of registers inserted between neighboring MACs.
MX-SAFE core contains 16 PUs in a 4$\times$4 configuration, where these PUs are pipelined for the systolic execution. 
Then, each PU contains 4$\times$4 MAC units which perform sub-matrix multiplication in parallel (Fig.~\ref{fig:MXSF_System}(b)).
Since each MAC unit has four multipliers, 16 inputs and 16 weights are provided at each cycle for parallel execution (Fig.~\ref{fig:MXSF_System}(c)).
As illustrated in Fig.~\ref{fig:MXSF_System}(d), we support taking both 1D and 2D MX blocks as input operands.
The 1D mapping assigns four different blocks to a single STA PU, whereas tiled mapping assigns two blocks. 
The tiled approach creates a minor accumulation overhead in the STA but advantageously allows for a dataflow design that does not require reshaping 2D tiles into a 1D format.

\subsection{MXSF-aware MAC Unit}
Fig.~\ref{fig:MXSF_System}(c) shows SAFE-MAC, a new MAC unit designed to support the \texttt{MXSF} format. 
Unlike conventional formats, \texttt{MXSF} represents two distinct numerical formats, \texttt{E2M5} and \texttt{E3M2}, within a single block. 
Consequently, each element must be decoded to its corresponding value prior to computation. 
The decoding logic (Fig.~\ref{fig:MXSF_System}(e)) interprets the data as the \texttt{E3M2} format if its 2\textsuperscript{nd} and 3\textsuperscript{rd} MSBs are both zero; otherwise, it is treated as the \texttt{E2M5} format.
The SAFE-MAC unit processes four input-weight pairs per cycle to compute one partial sum. 
To compute on any of two decoded \texttt{MXSF} formats, each data pair is first multiplied by using an \texttt{E4M5}-based multiplier (Mul.) that fully covers \texttt{E2M5} and \texttt{E3M2}.
The four multiplier outputs are then accumulated to calculate the partial sum using an \texttt{FP12\_E4M7} adder tree. 
We select \texttt{FP12} adders, which provide enough bit coverage, instead of \texttt{BF16} adders to minimize the hardware overhead coming from the adder tree in the SAFE-MAC.

\section{Experimental Results}\label{sec:eval}
\subsection{Experimental Setup}
\subsubsection{Software Setup}
We modified MX implementation from Microsoft\footnote{https://github.com/microsoft/microxcaling} to evaluate the effectiveness of our \texttt{MXSF} format on direct-cast inference and training.
The implementation includes \texttt{MXINT8}, \texttt{MXFP8\_E4M3}, block minifloat from BOOST~(i.e., \texttt{MXFP8\_E2M5}~\cite{boost}) and our \texttt{MXSF}. 
The BOOST uses a 48$\times$48 block, but for a fair comparison, we fixed all MX format's block sizes to 1$\times$64 for inference and 8$\times$8 for training (favorable choice for BOOST).
Utilizing the PyTorch framework, we copied pre-trained models from “timm” for a vision task and “Hugging Face~\cite{huggingface}” for LLM and VLM tasks in our experiments.

\subsubsection{Direct-cast Inference}
To compare accuracy of various MX formats on direct-cast inference (\texttt{FP32}$\xrightarrow{}$\texttt{MX}), we select an image classification task and language generation tasks. 
For the image classification, we tested ResNet-18~\cite{resnet_sb}, MobileNetV2~\cite{mobilenetv2} and MobileNetV4-based MNv4-Conv-S~\cite{mobilenetv4} as CNN benchmarks, while DeiT-B~\cite{deit} and Swin-S~\cite{swin} are used as Vision Transformer (ViT) benchmarks.
Also, EfficientViT-B3~\cite{efficientvit} and FastViT-S12~\cite{fastvit} are selected as CNN+ViT benchmarks.
As a dataset, a widely used ImageNet-1K validation dataset is selected.
For the language generation task, we used EleutherAI's lm-evaluation-harness~\cite{eval-harness} to conduct experiments on the LLaMA-3.2-3B~\cite{dubey2024llama}. 
We evaluate various data formats on three question-answering tasks, i.e., Boolean Question (BoolQ) and AI2 Reasoning Challenge (ARC-Easy and ARC-Challenge), as well as one language modeling task, WikiText2.

\subsubsection{Model Training}
We report the training quality of various MX formats on two different tasks: a vision task and a multimodal task. 
To compare the full-training performance, we selected three different models, i.e., ResNet-18, MobileNetV2, and DeiT-Tiny for ImageNet-1K classification. 
We used eight RTX 3090 GPUs with a batch size of 128 per GPU, trained for 100 epochs, and selected the AdamW optimizer with a learning rate of {1e-3} and a cosine decay scheduler.
To compare the training performance on a visual language model (VLM) using the MX format, we trained nanoVLM-222M~\cite{nanovlm} from Hugging Face using an open-source dataset.     
We set batch\_size = 256 and conducted training on an RTX6000 Blackwell GPU based on the detailed configuration from the original nanoVLM GitHub repository\footnote{https://github.com/huggingface/nanoVLM/tree/v0.1}.


\begin{table}[]
\caption{Direct-cast Inference Vision \& LLM Zero-shot Task\\ (MXFP8: \texttt{MXFP8\_E4M3}, BOOST: \texttt{MXFP8\_E2M5)}}
\label{tab:infer_result}
\centering\resizebox{\linewidth}{!}{
\begin{tabular}{|l|c|c|c|c|c|}
\hline
\multicolumn{6}{|c|}{Vision Task (dataset: ImageNet-1K)} \\ \hline
\textbf{Model}            & Baseline & MXINT8                                & MXFP8 & BOOST                                 & MXSF                                 \\ \hline \hline     
\multicolumn{6}{|c|}{CNN architecture} \\ \hline                                        
\textbf{ResNet-18} (Acc.) $\uparrow$         & 73.14    & {\color[HTML]{FE0000} \textbf{73.15}} & 72.27 & {\color[HTML]{6434FC} \textbf{73.14}} & 73.05                                 \\  \hline
\textbf{MobileNetV2} (Acc.) $\uparrow$      & 72.95    & {\color[HTML]{6434FC} \textbf{72.48}}                                 & 66.27 & {\color[HTML]{FE0000} \textbf{72.51}} & 72.46 \\ \hline

\textbf{MNv4-Conv-S} (Acc.) $\uparrow$      & 74.58    & 73.18                                 & 61.80 & {\color[HTML]{FE0000} \textbf{73.89}} & {\color[HTML]{6434FC} \textbf{73.60}} \\ \hline

\multicolumn{6}{|c|}{Vision transformer (ViT) architecture} \\ \hline
\textbf{DeiT-B} (Acc.) $\uparrow$      & 81.97    &  81.91                                 & 81.75 & {\color[HTML]{FE0000} \textbf{81.95}} & {\color[HTML]{6434FC} \textbf{81.94}} \\ \hline
\textbf{Swin-T} (Acc.) $\uparrow$      & 81.36    &  {\color[HTML]{6434FC} \textbf{81.35}}                                 & 81.09 & 81.32 & {\color[HTML]{FE0000} \textbf{81.36}} \\  \hline
\multicolumn{6}{|c|}{Hybrid ViT architecture} \\ \hline                                                   
\textbf{EfficientViT-B3} (Acc.) $\uparrow$ & 83.35    & 82.69                                 & 74.64 & {\color[HTML]{FE0000} \textbf{83.04}} & {\color[HTML]{6434FC} \textbf{82.79}} \\ \hline
\textbf{FastViT-S12} (Acc.) $\uparrow$     & 79.89    & 77.91                                 & 38.25 & {\color[HTML]{FE0000} \textbf{79.00}} & {\color[HTML]{6434FC} \textbf{78.84}} \\ \hline \hline
\multicolumn{6}{|c|}{LLM Task (Model: Llama3.2-3B)}                                                                              \\ \hline
\multicolumn{1}{|l|}{\textbf{Task}}           & \multicolumn{1}{l|}{Baseline} & \multicolumn{1}{l|}{MXINT8} & \multicolumn{1}{l|}{{MXFP8}} & \multicolumn{1}{l|}{{BOOST}} & \multicolumn{1}{l|}{MXSF} \\ \hline \hline                               
\multicolumn{1}{|l|}{\textbf{WikiText2} (PPL) $\downarrow$}      & \multicolumn{1}{c|}{9.27}              & \multicolumn{1}{c|}{9.30}            & \multicolumn{1}{c|}{9.36}           & \multicolumn{1}{c|}{\color[HTML]{FE0000} \textbf{9.28}}           & {\color[HTML]{FE0000} \textbf{9.28}}                               \\ \hline
\multicolumn{1}{|l|}{\textbf{BoolQ} (Acc.) $\uparrow$}          & \multicolumn{1}{c|}{72.94}             & \multicolumn{1}{c|}{72.42}           & \multicolumn{1}{c|}{70.46}          & \multicolumn{1}{c|}{\color[HTML]{6434FC} \textbf{72.78}}          & \color[HTML]{FE0000} \textbf{73.03}                              \\ \hline
\multicolumn{1}{|l|}{\textbf{ARC-e} (Acc.) $\uparrow$}      & \multicolumn{1}{c|}{74.49}             & \multicolumn{1}{c|}{\color[HTML]{FE0000} \textbf{74.71}}           & \multicolumn{1}{c|}{74.07}          & \multicolumn{1}{c|}{\color[HTML]{6434FC} \textbf{74.33}}          & 74.28                              \\ \hline
\multicolumn{1}{|l|}{\textbf{ARC-c} (Acc.) $\uparrow$} & \multicolumn{1}{c|}{42.66}             & \multicolumn{1}{c|}{\color[HTML]{FE0000} \textbf{42.41}}           & \multicolumn{1}{c|}{41.98}          & \multicolumn{1}{c|}{42.06}          & \color[HTML]{6434FC}\textbf{42.32}          \\ \hline

\multicolumn{6}{l}{\scriptsize * Best and second best are marked in {\color[HTML]{FE0000}red} and {\color[HTML]{6434FC}blue}, respectively.}\vspace{-2mm} \\

\end{tabular}}\vspace{-4mm}
\end{table}

\subsubsection{Hardware Design}
All hardware modules in Fig.~\ref{fig:MXSF_System} for \texttt{MXSF} are implemented in RTL and synthesized with 65nm CMOS technology running at 500MHz. 
For performance comparison, we implemented two baselines. 
The first baseline is a \texttt{BF16} baseline accelerator, which has the same STA configuration.
The second baseline uses a more efficient \texttt{MXFP4} format for the main STA core. 
However, a small 8$\times$8 \texttt{BF16} systolic array is included, as certain operations, such as $Q\cdot\!K^T$, require \texttt{BF16} to maintain accuracy. 
The area and power consumption for these accelerators are estimated using Synopsys Design Compiler.



\subsection{Accuracy on Direct-cast Inference}

Table~\ref{tab:infer_result} summarizes the direct-cast inference accuracy on the ImageNet classification and LLM zero-shot inference tasks.
In the domain of vision tasks, the \texttt{MXFP8\_E4M3} format exhibits a notable performance degradation relative to other 8-bit MX formats. 
This degradation is primarily attributed to its limited mantissa bit-width, which induces substantial quantization errors. 
The effect is particularly pronounced in lightweight models, where the resulting accuracy degradation can exceed 8\%. 
Conversely, the \texttt{MXINT8}, \texttt{BOOST}, and \texttt{MXSF} formats demonstrate superior robustness by effectively mitigating performance loss, constraining accuracy degradation to within 1.1\%. 
Moreover, in models where values satisfying $S_e-e_x \geq 3$ are prevalent, i.e., hybrid ViTs, both \texttt{BOOST} and \texttt{MXSF} have been observed to outperform \texttt{MXINT8}, yielding higher final accuracy.
For LLM tasks, other data formats still show slightly better performance compared to \texttt{MXFP8\_E4M3}.
Based on these experiments, we show that the \texttt{MXSF} achieves similar accuracy as \texttt{BF16} for the direct-cast inference.

\begin{table}[t]
\caption{Accuracy on Training Vision Tasks (dataset: ImageNet-1K)}
\label{tab:training_result}
\centering
\begin{tabular}{|cccccc|}
\hline
\multicolumn{1}{|c|}{\textbf{Model}}               & \multicolumn{1}{c|}{Baseline} & \multicolumn{1}{c|}{MXINT8}                                 & \multicolumn{1}{c|}{MXFP8}                            & \multicolumn{1}{c|}{BOOST}                                & MXSF                                 \\ \hline \hline

\multicolumn{1}{|c|}{\textbf{ResNet-18}}      & \multicolumn{1}{c|}{69.0}            & \multicolumn{1}{c|}{56.9}                                 & \multicolumn{1}{c|}{{\color[HTML]{6434FC} \textbf{65.3}}} & \multicolumn{1}{c|}{62.2}                                 & {\color[HTML]{FE0000} \textbf{68.4}} \\ \hline
    \multicolumn{1}{|c|}{\textbf{MobileNetV2}}     & \multicolumn{1}{c|}{69.9}            & \multicolumn{1}{c|}{31.1}                                 & \multicolumn{1}{c|}{{\color[HTML]{6434FC} \textbf{63.2}}} & \multicolumn{1}{c|}{46.2}                                 & {\color[HTML]{FE0000} \textbf{69.5}} \\
    \hline 
    \multicolumn{1}{|c|}{\textbf{DeiT-Tiny}}     & \multicolumn{1}{c|}{67.2}            & \multicolumn{1}{c|}{50.3}                                 & \multicolumn{1}{c|}{{\color[HTML]{6434FC} \textbf{65.4}}} & \multicolumn{1}{c|}{63.0}                                 & {\color[HTML]{FE0000} \textbf{66.7}} \\
    \hline 
    \end{tabular}\vspace{-2mm}
    \end{table}
    
    \begin{figure}[t]
        \centering
        \includegraphics[width=0.8\linewidth]{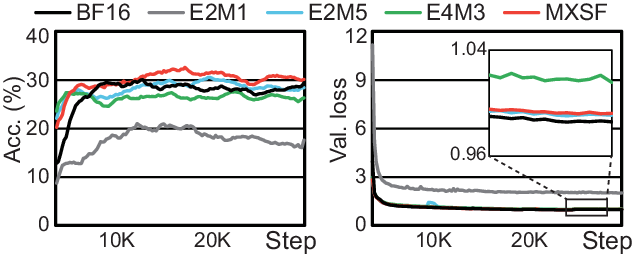}\vspace{-3mm}
        \caption{Training curve for nanoVLM-222M using various MX formats: accuracy (left) and validation loss (right) on the MMStar~\cite{mmstar} benchmark.}
        \label{fig:nanovlm_result}\vspace{-4mm}
    \end{figure}

    \begin{table}[t]
    \centering
    \caption{Area and Power Breakdowns of MX-SAFE accelerator}
    \label{tab:hardware_breakdown}
    \resizebox{0.9\linewidth}{!}{%
    {\scriptsize
    \begin{tabular}{|l|r|r|}
\hline
  \textbf{Ours (65nm, 500MHz)}                   & \multicolumn{1}{c|}{\textbf{Area ($\mu$m\textsuperscript{2})}} & \multicolumn{1}{c|}{\textbf{Power (mW)}} \\ \hline \hline
\textbf{MX-SAFE Core}       & 1,784,575.55 (87.90\%)                               & 308.845 (84.18\%)               \\ \hline
\textbf{MXSF Converter}            & 14,925.96 (0.74\%)                                 & 7.09 (1.93\%)                 \\ \hline
\textbf{Act. Function Unit}         & 10,969.92 (0.54\%)                                 & 2.61 (0.71\%)                 \\ \hline
\textbf{Softmax}              & 181,606.47 (8.94\%)                                & 24.12 (6.57\%)                \\ \hline
\textbf{Norm. Unit}              & 38,267.18 (1.88\%)                                & 24.21 (6.60\%)                \\ \hline

\end{tabular}%
}
}
\end{table}

\subsection{Accuracy on Model Training}
Other than inference, we also tested the stability of the \texttt{MXSF} format on training tasks.
Table~\ref{tab:training_result} summarizes the ImageNet-1K training accuracy of full-training scenarios. 
Since small gradients are important throughout the entire training process, both the \texttt{MXFP8\_E4M3} and \texttt{MXSF} formats show excellent performance in preventing underflow.
Furthermore, by allocating more mantissa bits to values near the shared exponent, the \texttt{MXSF} format improves its resilience to quantization error, resulting in an average of 3.6\% higher training accuracy than \texttt{MXFP8\_E4M3} in classification tasks.

As shown in Fig.~\ref{fig:nanovlm_result}, the nanoVLM training task involves fine-tuning a combination of a pre-trained vision encoder (SigLIP-B~\cite{siglip}) and a small language model (SmolLM2-135M~\cite{smollm2}) for the multimodal objective. 
Unlike the full-training case, fine-tuning is performed on pre-trained weights making significantly small gradients to appear less, which explains why \texttt{FP8\_E2M5} achieves high training accuracy.
Nevertheless, even in this scenario, \texttt{MXSF} is observed to deliver slightly better accuracy compared to other MX formats.

\begin{figure}[t]
    \centering
    \includegraphics[width=1\linewidth]{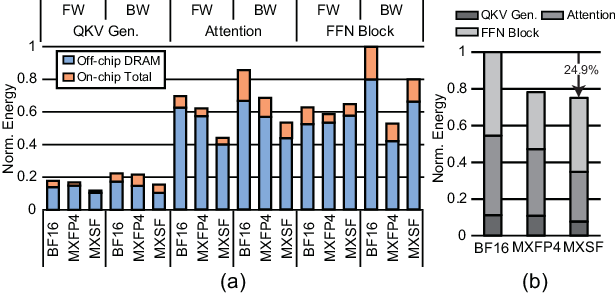}\vspace{-3mm}
    \caption{Energy consumption of DeiT-Tiny during a single-batch training. (a) Breakdowns per encoder block and (b) overall system energy consumption.}
    \label{fig:energy}\vspace{-4mm}
\end{figure}

\subsection{Hardware Analysis}

Based on the area and power consumption in Table~\ref{tab:hardware_breakdown}, we performed a detailed hardware analysis using DeiT-Tiny model on ImageNet.
Table~\ref{tab:hardware_breakdown} shows that the MX-SAFE core is the dominant hardware component, consuming the majority of both the total area (87.68\%) and power (84.18\%) of the MX-SAFE accelerator.
As shown in Fig.~\ref{fig:energy}, the total energy consumption is primarily driven by the off-chip memory access, which accounts for 83.93\%. 
The remaining energy is consumed by the on-chip memory access and the core, contributing 14.92\% and 1.15\%, respectively. 
For the energy estimation, we utilized formulas from BitMoD~\cite{bitmod} to model the off-chip and on-chip memory accesses.
Note that \texttt{MXFP4} consumes more energy on Attention layers since it uses \texttt{BF16} for $Q\cdot\!K^T$ and $Attn\cdot\!V$.
In return, the proposed \texttt{MXSF} format demonstrates a 24.92\% reduction in total energy consumption compared to the \texttt{BF16} baseline.
Furthermore, a comparison with the \texttt{MXFP4} hardware, which includes a small \texttt{BF16} systolic array, shows that our design is 4.07\% more energy-efficient, underscoring the accuracy and the energy efficiency of our \texttt{MXSF} format. \vspace{-2mm}

\section{Conclusion}
In this paper, we proposed MX-SAFE (\texttt{MXSF}) format, which safely supports both inference and training by allowing a large mantissa mode (\texttt{E2M5}) and a high dynamic range mode (\texttt{E3M2}) within the same block.
\texttt{MXSF} proves to be an outstanding substitute for \texttt{BF16} in both training and inference. 
It results in a negligible accuracy drop compared to BF16—only 0.17\% on average for direct cast inference and under 0.5\% for training—distinguishing it from other MX formats.
Furthermore, to optimize training efficiency, we minimized both the memory usage during training and the quantization process by employing tile-based blocks instead of the 1D blocks used in the existing MX format.
Overall, we present MX-SAFE accelerator, i.e., a training-inference accelerator based on the \texttt{MXSF} format. 
Our \texttt{MXSF}-based hardware accelerator achieves accuracy similar to \texttt{BF16} baseline while reducing total energy consumption, including off-chip memory accesses, by 24.9\% in the DeiT-Tiny training task.
\bibliographystyle{IEEEtran}
\bibliography{reference}

\vspace{12pt}
\end{document}